# Joint operation in public key cryptography

Dragan Vidakovic[1], Olivera Nikolic[1], Dusko Parezanovic[2] and Jelena Kaljevic[1]
[1] Faculty of Business Valjevo, Singidunum University Belgrade, Serbia

[2]Club of young math. and computer scientists Infomat Ivanjica, Serbia
dragan.vidakovic@open.telekom.rs
{onikolic, jkaljevic}@singidunum.ac.rs
infomat@open.telekom.rs

***Abstract.*** **We believe that there is no real data protection without our own tools. Therefore, our permanent aim is to have more of our own codes. In order to achieve that, it is necessary that a lot of young researchers become interested in cryptography. We believe that the encoding of cryptographic algorithms is an important step in that direction, and it is the main reason why in this paper we present a software implementation of finding the inverse element, the operation which is essentially related to both ECC (Elliptic Curve Cryptography) and the RSA schemes of digital signature.**

## I. INTRODUCTION

It has already been mentioned that we believe the best protection is achieved by developing our own software. However, this process requires a lot of knowledge, skills, patience and great responsibilities [7], provided that the undisputed cryptic of cryptography itself was previously overcome and that there are courage and willingness to enter more deeply into the matter.

In this paper we want to show that it is possible to implement the inverse element without any software-hardware facilities (in the arithmetic of large numbers), which is a very important operation in the process of realization of both leading schemes of a public key – ECC and RSA [1] [4] [6].

## II. TASK AND AIM

In the arguments for and against in a trial of strength of ECC and RSA, the simple fact that they are performed by the same tools made for operations with large numbers, is usually overlooked. Mathematical bases of RSA and ECC are completely different [1] [5], but they need the same operations: addition, subtraction, multiplication, division, finding the remainder, calculating d from the equation $e*d \equiv 1 \pmod p$ for fixed values of e and p, and more joint auxiliary operations needed for the realization of a digital signature.

When it comes to joint operations, we have opted for finding the inverse element. Therefore, we will present a brief review of the RSA and ECC and point out those steps in algorithms in which it is necessary to calculate the inverse element.

### *A. Algorithm Key Generation for the RSA Signature Scheme*

As we can see, for the given public key e and the number $\phi$, d should be calculated from the equation in order to get e private key – the most important element of the RSA digital signature scheme, or, in other words, we should find the inverse element of the element e.

Sumary: each entity creates an RSA public key and a corresponding private key. Each entity A should do the following [1]:

1. Generate two large distinct random primes p and q, each roughly the same size (see x11.3.2).
2. Compute $n = pq$ and $\phi= (p - 1)(q - 1)$.
3. Select a random integer e, $1 < e < \phi$ such that $\gcd(e, \phi) = 1$.
4. Use the extended Euclidean algorithm ([2]) to compute the unique integer d, $1 < d < \phi$, such that $ed \equiv 1 \pmod \phi$)
5. A's public key is (n; e); A's private key is d

As we can see for a given public key e and modulo $\phi$, we have to find inverse element d- secret key- for RSA digital signature. So, finding the inverse element is very important stage the RSA signature scheme.

### *B. The shortest of the ECC*

Efficient and secure public-key cryptosystems are essential in today's age of rapidly growing Internet communications. Elliptic curve scalar multiplication in particular, which refers to the operation of multiplying a large integer by a point on an elliptic curve, is crucial for both data encryption technology as well as testing the security of cryptographic systems.[5]

In reality it's very hard to find k (large number) that k*P=Q where P and Q are given points on an Eliptic curve.(P, Q are public keys, and k is private or secret key)

Practicaly we have to solve *P+P and P+Q to* get thrid point R=P+P or R=P+Q.

To put it simply, in the field, say $F_p$ , where p is prime, for two points $P(x_p,y_p)$, $Q(x_q,y_q)$ on an elliptic curve, using chord-tangent-rule, to give the third elipting curve point $R(x_r,y_r)$.

Coordinates of the third point are calculated by the following formulas:

$x_r = ((y_q - y_p)_/ (x_q - x_p))^2 - x_p - x_q$ ,

$y_r = ((y_q - y_p)_/ (x_q - x_p))(x_p - x_r) - y_p$

It is obvious that it is very important to find $(x_q - x_p)^{-1}$

## III. CODE FOR FINDING THE INVERSE ELEMENT

In the previous paragraph, we pointed out the place and significance of finding the inverse element of a fixed element in the field Fp, where p is a prime. Since our goals are to avoid the general story and to develop the software that can operate in real conditions (with large numbers), in this paragraph we will present the code by which we can solve the equation $ed \equiv 1 \pmod{\phi}$ for (arbitrarily) large fixed number $\phi$ and the fixed e.

### *A. The Code for Calculating the Inverse Element – Delphi 7 Console Application*

In order to calculate the inverse element, it is necessary to encode Binary extended gcd algorithm. For the fixed inputs e and $\phi$, wwe will get the output d, and this will be the solution of the modular equation above.

```
Program for_inverse;
{$APPTYPE CONSOLE}
uses SysUtils, Unit22;
label 1,2;
var a,b,c,g,u,v,a1,b1,d:array[0..nn1] of integer;
gcd,w:array[0..nn] of integer;
i,s1,s2,i1,p1,c1:integer;
ff:file of integer;
 x,y,k:array[0..nn] of integer;
k1:longint;
begin
s1:=0; s2:=0;
writeln('calculating the inverse element');
 assign(ff,'brojfi.dat'); rewrite(ff);
   {Example module (10111)-binary, (23)-decimal}
    x[0]:=1;x[1]:=1;x[2]:=1;x[3]:=0;x[4]:=1;
write(ff,x[0]);write(ff,x[1]);write(ff,x[2]);write(ff,x[3
]);write(ff,x[4]);
reset(ff); i:=0;
while not eof(ff) do
begin
read(ff,x[i]); i:=i+1;
end;
assign(ff,'javni.dat'); rewrite(ff);
{Example elements (100)-binary, 4 decimal}
y[0]:=0;y[1]:=0;y[2]:=1;y[3]:=0;
write(ff,y[0]);write(ff,y[1]);write(ff,y[2]);write(ff,y[3
]);
reset(ff); i:=0;
while not eof(ff) do
begin
read(ff,y[i]); i:=i+1;
end;
s1:=0; s1:=0;s2:=0;
  for i:=0 to nn do
 begin

g[i]:=0;u[i]:=0;v[i]:=0;a[i]:=0;b[i]:=0;c[i]:=0;d[i]:=0;
w[i]:=0;
  end;
   g[0]:=1;p1:=0;
  while ((x[0]=0) and (y[0]=0)) do
  begin
   for i:=1 to nn do  begin
   x[i-1]:=x[i]; y[i-1]:=y[i];end;
    for i:=nn-1 downto 0 do g[i+1]:=g[i];
  g[p1]:=0;p1:=p1+1; end;
    for i:=0 to nn do begin
  u[i]:=x[i]; v[i]:=y[i];end;
    a[0]:=1;b[0]:=0;c[0]:=0;d[0]:=1;
    1:   while u[0]=0 do
   begin
         for i:=1 to nn do u[i-1]:=u[i];
              if ((a[0]=0) and (b[0]=0)) then begin
                  for i:=1 to nn do a[i-1]:=a[i];
                   for i:=1 to nn do b[i-1]:=b[i];  end
                                 else  begin
                                 saberi(a,y,w);
                     for i:=1 to nn do w[i-1]:=w[i];
                     for i:=0 to nn do a[i]:=w[i];
                   oduzmi(b,x,w);
                     for i:=1 to nn do w[i-1]:=w[i];
                     for i:=0 to nn do b[i]:=w[i]; end;end;
     while v[0]=0 do begin
          for i:=1 to nn do v[i-1]:=v[i];
           if ((c[0]=0) and (d[0]=0)) then begin
              for i:=1 to nn do c[i-1]:=c[i];
               for i:=1 to nn do d[i-1]:=d[i]; end
                      else  begin
                    saberi(c,y,w);
                  for i:=1 to nn do w[i-1]:=w[i];
                  for i:=0 to nn do c[i]:=w[i];
                  oduzmi(d,x,w);
                  for i:=1 to nn do w[i-1]:=w[i];
                  for i:=0 to nn do d[i]:=w[i]; end;
veci1(v,s1);   end;
   i:=nn;veci1(u,s1);
   while u[i]=v[i] do i:=i-1;
   if i<0 then i:=i+1;
```

```
  if u[i]>=v[i] then  begin
     oduzmi(u,v,w);
    for i1:=0 to nn do u[i1]:=w[i1];
    oduzmi(a,c,w);
    for i1:=0 to nn do a[i1]:=w[i1];
    oduzmi(b,d,w);
    for i1:=0 to nn do b[i1]:=w[i1]; end
           else begin
             oduzmi(v,u,w);
    for i1:=0 to nn do v[i1]:=w[i1];
             oduzmi(c,a,w);
             for i1:=0 to nn do c[i1]:=w[i1];
      oduzmi(d,b,w);
    for i1:=0 to nn do d[i1]:=w[i1]; end;
          s1:=0;veci1(u,s1);
      if s1<>0 then goto 1;
       for i1:=0 to nn do a1[i1]:=c[i1];
            for i1:=0 to nn do k[i1]:=d[i1];
  s1:=0;   veci1(g,s1);   s2:=0;   veci1(v,s2);
if ((s1=1) and (g[0]=1) and (s2=1) and (v[0]=1)) then
begin
  s1:=0;
   veci1(k,s1);
  if s1<0 then begin
  for i:=0 to nn do
    k[i]:=abs(k[i]);
       oduzmi(x,k,k);
      end;
  dokle(y,s2);
  dokle(k,s2);
  assign(ff,'tajni.dat');
  rewrite(ff);
 for i:=0 to s2 do
    write(ff,k[i]);
    reset(ff);    i:=0;
    while not eof(ff) do
    begin
      read(ff,k[i]);       i:=i+1;
      end;
      dokle(k,s2);
      for i:=s2 downto 0 do
      write(k[i]);   writeln;
  writeln('found a secret key for RSA-The inverse
element for ECC');
  readln;
    end
  else
  writeln('no inverse element,
gcd(module,elements)<>1');
  readln;
    end.
```

### *B. The unit that serves the program above (III.A)*

The program above, together with the unit Unit22, can be run as a Delphi 7 console application in order to do the testing and present the result of the two examples on both schemes of a public key.

Besides the basic operations in arithmetic of large numbers, the unit also contains some auxiliary functions.

```
unit Unit22;
 interface
{  for larger numbers set larger nn, nn1 }
const nn=100;
const nn1=100;
procedure saberi(x,y:array of integer;var w:array of
integer);
procedure oduzmi(x,y:array of integer;var w:array of
integer);
procedure veci1(var x:array of integer;var s:integer);
procedure  dokle(var a:array of integer;var
s1:integer);
implementation
procedure saberi(x,y:array of integer;var w:array of
integer);
var c1,i1,s1,s2:integer;
begin
s1:=0;s2:=0;veci1(x,s1);veci1(y,s2);
if ((s1>=0) and (s2>=0)) then begin
c1:=0;
for i1:=0 to nn1 do begin
  w[i1]:=(x[i1]+y[i1]+c1) mod 2;
  if (x[i1]+y[i1]+c1)<2 then c1:=0
              else c1:=1;end;
  w[nn1]:=c1; end
    else
  if ((s1>=0) and (s2<0)) then begin
  for i1:=0 to nn1 do y[i1]:=abs(y[i1]);
    oduzmi(x,y,w);end
        else
       if ((s1<0) and (s2>=0)) then begin
        for i1:=0 to nn1 do x[i1]:=abs(x[i1]);
         oduzmi(y,x,w);end
            else if ((s1<0) and (s2<0)) then begin
          for i1:=0 to nn1 do x[i1]:=abs(x[i1]);
            for i1:=0 to nn1 do y[i1]:=abs(y[i1]);
            saberi(x,y,w);
            for i1:=0 to nn1 do w[i1]:=-w[i1]; end; end;
  procedure oduzmi(x,y:array of integer;var w:array
of integer);
  label 1;
 var i1,c1,k,s1,s2:integer;
 begin
 s1:=0;s2:=0;veci1(x,s1);veci1(y,s2);
  if ((s1>=0) and (s2>=0)) then begin
  k:=0;
1: c1:=0;
 for i1:=0 to nn1 do begin
  w[i1]:=abs(x[i1]-y[i1]+c1) mod 2;
 if (x[i1]-y[i1]+c1 )>=0 then c1:=0
```

```
         else c1:=-1; end;
    if k=1 then
   c1:=0;
   if c1=-1 then begin
    for i1:=0 to nn do x[i1]:=0;
    for i1:=0 to nn do y[i1]:=w[i1];
     c1:=0;k:=1;goto 1;end;
    if k=1 then
    for i1:=0 to nn do w[i1]:=-w[i1];end
    else if ((s1>=0) and (s2<0)) then begin
       for i1:=0 to nn do y[i1]:=abs(y[i1]);
saberi(x,y,w);
        end
      else if ((s1<0) and (s2>=0)) then begin
          for i1:=0 to nn1 do
x[i1]:=abs(x[i1]);saberi(x,y,w);
         for i1:=0 to nn1 do w[i1]:=-w[i1];end
            else if ((s1<0) and (s2<0)) then begin
               for i1:=0 to nn1 do x[i1]:=abs(x[i1]);
                  for i1:=0 to nn1 do y[i1]:=abs(y[i1]);
                  oduzmi(y,x,w); end; end;
  procedure veci1(var x:array of integer;var s:integer);
      var i1:integer;
      begin
      s:=0;
      for i1:=0 to nn do s:=s+x[i1]; end;
procedure  dokle(var a:array of integer;var
s1:integer);
begin
s1:=nn;
while a[s1]=0 do s1:=s1-1;
end;
 procedure koji(var u,v:array of integer;var l:integer);
 var i:integer;
 begin
 l:=0; i:=nn;
  while u[i]=v[i] do i:=i-1;
   if i<0 then l:=0
   else
   if u[i]>v[i] then l:=1
   else l:=-1;
   end;
end.
```

## IV. EXAMPLES

Here, we will give two examples to illustrate the use of inverse element in RSA and ECC.

### *A. RSA example*

In [2] we can find solutions of the equation ed ≡1 (mod ϕ) for fixed values e, and ϕ**.** More precisely, for the public key e, and a module ϕ we can calculate the secret key d, element required for RSA scheme.

If ϕ-module**:**

10000000000000000000000000000000000000000000000000
000000000000000000000000000000000000000000000000000000
000000000000000000000000000000000000000000000000000000
000000000000000000000000000000000000000000000000000000
000000000000000000000000000000000000000000010000000000
000000000000000000000000000000000000000000000000000000
000000000000000000000000000000000000000000000000000000
000000000100000000000000000000000000100000000000000000
000000000000000000000000000000000100000000000000000000
000000000000000000110100011110100000000000000000000000
000000000000000000000000000000000000000000000000000000
000000000000000000000000000000000000000000000000010000
000000000000000000000010000000000000000000000000000000
000000000000000000010000000000000000000000000000000000
000011010001111010000000000000000000000001000000000000
000000000000000010000000000000000000000100000000000000
000000000000100000000000110100011110000000000000000110
100011110000000000000000000000000000000000000000111100
000100000000000000000000000000000101010000111010011100
0.

e-public key: 111

Then d-RSA secret key:

10010010010010010010010010010010010010010010010010
010010010010010010010010010010010010010010010010010010
010010010010010010010010010010010010010010010010010010
010010010010010010010010010010010010010010010010010010
010010010010010010010010010010010010010100100100100100
100100100100100100100100100100100100100100100100100100
100100100100100100100100100100100100100100100100100100
100100101001001001001001001001001001001101101101101101
101101101101101101101101101101110010010010010010010010
010010010010010011001110010000100100100100100100100100
100100100100100100100100100100100100100100100100100100
100100100100100100100100100100100100100100100100110110
110110110110110110111001001001001001001001001001001001
001001001001001001011011011011011011011011011011011011
011111001011010100100100100100100100100101001001001001
001001001001001011011011011011011011011100000000000000
000000000000100100100101100000100010010010010010011001
110010000000000000000000000000000000000000000001000100
1010010010010010010010010010010011110010100010111111.

### *B. ECC example*

Let p=23. Consider an elliptic curve $y^2 = x^3 +x+4$ defined over $F_{23}$ $(E(F_{23}))$ [4]. Using code in III. A we can find coordinates of the point R=P+Q.

If P(4,7) and Q(13,11), based on the relation of II. B. we can calculate the x coordinate of the point R.

$x_r =((11-7)/(13-4))^2 -4-13 \pmod{23}$

Using code (III. A.) we can calculate $(13-4)^{-1} =9^{-1}$ .

*Output on the screen:*

*calculating the inverse element*

*10010*

*found a secret key for RSA-The inverse element for ECC.*

We can see that 10010 (base 2) is 18 (base 10). Indeed: 9*18=162, 162 mod 23=1, so $9^{-1}$ =18 is the inverse element. For further computation, the other necessary operations are located in Unit22 (subtraction and addition-In Serbian language *saberi, oduzmi)* and in [2], *mnozi, ostatatk (* multiplication, and the remainder in English) and many auxiliary functions, which we will show in the forthcoming paper.

## V. CONCLUSION

We believe that each country must stimulate young people's interest in cryptography, because we doubt that our secret data can be protected using someone else's software [3].

Of course, it is very difficult to develop our own protection mechanisms, but we think it is far better to protect data using our own mechanisms first, and then, thus modified, leave them to someone else's software, than to allow the original data be protected by somebody else's mechanisms, which is a logical nonsense.

That is the reason why we always insist on more our own softwares and a greater interest in cryptography, which seems itself (in case it wasn't brought closer to a reader) pretty cryptic and bouncing. So, this work is primarily addressed to young researches as an incentive to try to develop their own tools for data protection. Those tools do not have to be flawless, they may be far below the level of the tools found on the market. However, they should be good enough for the beginning of a hard work that would lead researches to some great commercial solutions.